\documentclass[reprint, amssymb, amsmath, aip,jcp]{revtex4-1}

\usepackage{graphicx}
\usepackage{dcolumn}
\usepackage{bm}

\newcommand{\bra}[1]{\langle #1|}
\newcommand{\ket}[1]{|#1\rangle}
\newcommand{\braket}[2]{\langle #1|#2\rangle}
\usepackage{color}

\begin{document}


\title{
Phase Space Approach to Laser-driven Electronic Wavepacket Propagation}
\thanks{
The following article has been submitted to the Journal of Chemical Physics. 
After it is published, it will be found at http://jcp.aip.org/.
}

\author{Norio Takemoto}
\author{Asaf Shimshovitz}
\author{David J. Tannor}
\affiliation{Department of Chemical Physics, 
   Weizmann Institute of Science, 76100 Rehovot, Israel}

\date{\today}

\begin{abstract}
We propose a phase space method to propagate a quantum 
wavepacket driven by a strong external field. 
The method employs the so-called biorthogonal von Neumann basis 
recently introduced for the calculation of the energy 
eigenstates of time-independent quantum systems
[A. Shimshovitz and D.J. Tannor, arXiv:1201.2299v1]. 
While the individual elements in this basis set are time-independent, 
a small subset is chosen in a time-dependent manner 
to adapt to the evolution of the wavepacket in phase space. 
We demonstrate the accuracy and efficiency of the present propagation 
method by calculating the electronic wavepacket in a
one-dimensional soft-core atom interacting with a superposition of
an intense, few-cycle, near-infrared laser pulse
and an attosecond extreme-ultraviolet laser pulse. 
\end{abstract}

\maketitle


With the emergence of attosecond laser 
technology, \cite{naturePhotonics4_822}
there is the fascinating prospect of
observing and controlling the correlated dynamics 
of multiple electrons on its natural time scale of 
ten to one hundred attoseconds. \cite{revModPhys81_163}
In order to unravel the complex and 
sometimes counter-intuitive \cite{prl105_203004,prl107_143004} 
quantum dynamics from the experimental data, 
and to develop theories
that reproduce the essence of the dynamics \cite{prl71_1994},
an accurate and efficient numerical method to
simulate the multi-electron wavepacket dynamics is indispensable.

However, accurate simulation of the electronic dynamics 
in a high-intensity laser field is a challenging task:
the electronic wavepacket is dispersed by the laser field
over a wide region of coordinate space while retaining high
momentum near the atomic nuclei. 
Straightforward representation of the wavefunction
on a equally-spaced coordinate grid [i.e., a Fourier grid (FG)] requires 
a large range with a small interval between points.
Simulation on such a large grid quickly becomes 
prohibitive as the number of degrees of freedom (DOF) increases.
Even with sophisticated treatments such as 
multi-configuration time-dependent Hartree-Fock 
(MCTDHF), \cite{newJPhys10_025019}
simulation of ionization dynamics 
have been limited to small systems such as 
the helium atom and the hydrogen molecule. \cite{jPhysB36_L393,
jcp113_8953, laserPhys13_1064,jcp128_184102,
prl96_073004,physRevA83_063416}

In this article, we present a new approach to solving the
time-dependent Schr\"{o}dinger equation (TDSE), based on
a phase space perspective. The resultant propagation method 
is simple, accurate, stable, and efficient. As a first
demonstration, we simulate the electronic wavepacket of a
one-dimensional (1D) model atom in the combined fields of 
a high-intensity near-infrared (NIR) laser pulse and an
attosecond extreme-ultraviolet (XUV) laser pulse.

In our approach,
we utilize the localized nature of phase space Gaussians
to prune the basis.
However, a basis set of phase-space localized states is perforce
non-orthogonal \cite{Low1985},
and this has created difficulties in 
previous attempts \cite{jcp113_10028,chemPhys304_103,jcp115_1158,jcp118_2606,
jcp124_204101,jcp118_6720}
to truncate the basis 
to cover only the phase space region of 
dynamics. 
This long standing problem was solved recently,
by the so-called periodic von Neumann (pvN) 
basis. \cite{newJPhys11_105052,arXiv1201_2299v1}
The pvN basis is generated from a set of Gaussians whose
centers form a finite lattice in phase space. By imposing
periodic boundary conditions on these Gaussians, 
the pvN basis becomes 
formally equivalent to the simple and accurate 
FG representation. \cite{Kosloff1993}
Then, its biorthogonal basis, the biorthogonal von Neumann (bvN) basis, 
can be used for a compact representation of a quantum state.
This framework 
was successfully applied to the calculation 
of quantum energy eigenstates. \cite{arXiv1201_2299v1}
Here, the framework is extended for solving the TDSE.
Our strategy is to keep the individual elements in the basis set 
time-independent
but to 
truncate the bvN basis in a time-dependent manner. 
This avoids the problem of a moving basis \cite{chemPhys304_103} 
that can become over-complete as time elapses and can become unstable.
In contrast,
here we obtain 
a stable set of linear ordinary differential equations (ODEs) 
for the expansion coefficients in the truncated bvN basis.


Before explaining the pvN and bvN bases, 
we review the formalism of the FG basis to establish notation.
We write the Fourier pseudo-spectral and spectral
bases \cite{jcp91_3571,jCompPhys47_412,jPhysChem92_2087}
as $\{\ket{\theta_m}\}_{m=1,\dots,N}$ and  $\{\ket{\phi_m}\}_{m=1,\dots,N}$,
respectively. 
These bases are orthonormal and
span the same $N$-D Hilbert space
denoted here as $\mathcal{H}$,
resolving the identity in $\mathcal{H}$ as
\begin{equation}
1_{\mathcal{H}} 
= \sum_{m=1}^N \ket{\phi_m}  \bra{\phi_m} 
= \sum_{m=1}^N \ket{\theta_m}  \bra{\theta_m} .
\end{equation}
The bases $\{\ket{\theta_m}\}$ and $\{\ket{\phi_m}\}$ 
are localized 
at the FG points $\{x_m\}_{m=1,\dots,N}$
in the position space and 
$\{p_m\}_{m=1,\dots,N}$ in the momentum space, respectively. 
For any quantum state $\ket{\Psi} \in \mathcal{H}$,
$\braket{\theta_m}{\Psi} = \braket{x_m}{\Psi}\sqrt{\Delta x}$
and 
$\braket{\phi_m}{\Psi} = \braket{p_m}{\Psi}\sqrt{\Delta p}$,
where $\Delta x$ and $\Delta p$ are the grid intervals 
in position and momentum spaces, respectively. 

The pvN basis \cite{newJPhys11_105052,arXiv1201_2299v1}
$\{\ket{\tilde{g}_j}\}_{j=1,\dots,N}$ is defined as 
\begin{equation}
\ket{\tilde{g}_j} 
= \sum_{m=1}^N \ket{\theta_m} \braket{x_m}{g_j} \sqrt{\Delta x},
\end{equation}
where $\{\ket{g_j}\}_{j=1,\dots,N}$ are the phase space Gaussians,
\begin{equation}
\begin{split}
\braket{x}{g_j} 
=& \left(\frac{\gamma}{\pi}\right)^{1/4}
  \exp\left(-\frac{\gamma}{2} (x-q_j)^2 + \frac{i}{\hbar} p_j (x-q_j)
\right.\\& \left. \phantom{nnnnnnnnnnnnnnnnnnn}
            +\frac{i}{2\hbar} p_j q_j\right),
\end{split}
\end{equation}
whose centers $\{(q_j, p_j)\}_{j=1,\dots,N}$ 
constitute a finite lattice in the phase space
with the unit cell of area $2\pi \hbar$ 
and the momentum-to-position aspect ratio $\hbar\gamma$. 
Note that the number of Gaussians $N$ is the same as the 
number of FG points used.

The bvN basis \cite{arXiv1201_2299v1}
$\{\ket{\tilde{b}_j}\}_{j=1,\dots,N}$ 
is defined to be biorthogonal (dual) to the pvN basis, i.e.,
$\braket{\tilde{b}_l}{\tilde{g}_j}= \delta_{lj}$. 
This gives 
\begin{equation}
\ket{\tilde{b}_j} 
= \sum_{l=1}^N \ket{\tilde{g}_l} (S^{-1})_{lj},
\end{equation}
where $S^{-1}$ is the inverse of the overlap matrix 
$S_{lj}
  = \braket{\tilde{g}_l}{\tilde{g}_j} 
  = \Delta x \sum_{m=1}^N \braket{g_l}{x_m}\braket{x_m}{g_j} $ 
of the pvN basis. 
The pvN and bvN bases 
span the same Hilbert space $\mathcal{H}$
as $\{\ket{\phi_m}\}$ and $\{\ket{\theta_m}\}$,
resolving the identity as
\begin{equation}
1_{\mathcal{H}} 
= \sum_{j=1}^N \ket{\tilde{b}_j}\bra{\tilde{g}_j}
= \sum_{j=1}^N \ket{\tilde{g}_j}\bra{\tilde{b}_j}.
\end{equation}
Thus, any wavepacket $\ket{\Psi(t)} \in \mathcal{H}$ 
can be represented as 
\begin{equation}
\ket{\Psi(t)}
= \sum_{j=1}^N \ket{\tilde{b}_j}\braket{\tilde{g}_j}{\Psi(t)}.
\end{equation}
Due to the localized nature of the Gaussians $\{\ket{g_j}\}$,
the magnitude of $\braket{\tilde{g}_j}{\Psi(t)}$ can be extremely small
if the corresponding classical system can not reach
the phase space region around $(q_j, p_j)$. 
Defining a set $\mathcal{A}$ 
such that $|\braket{\tilde{g}_j}{\Psi(t)}|$ is negligible
if $j\not\in \mathcal{A}$, we can approximate the 
wavepacket by a subset of the bvN basis as
\begin{equation}\label{eq: Psi approx}
\ket{\Psi(t)} 
\approx \sum_{j \in \mathcal{A}} \ket{\tilde{b}_j} c_j(t),
\end{equation}
where $c_j(t) := \braket{\tilde{g}_j}{\Psi(t)}$.
Note that the set $\mathcal{A}$ of active indices can be 
changed in time in order to keep the number $N_{\mathcal{A}}$ of 
elements in $\mathcal{A}$ small at all time.

By substituting eq. (\ref{eq: Psi approx}) to the TDSE,
we obtain a set of linear ODEs
for the active bvN coefficients $\{c_j\}_{j\in \mathcal{A}}$,
\begin{equation}\label{eq: TDSE} 
\frac{d c_j}{dt}
= -\frac{i}{\hbar} \sum_{l \in \mathcal{A}} \sum_{m\in \mathcal{A}} 
  (\Omega^{-1})_{j l} 
  \bra{\tilde{b}_l} H(t) \ket{\tilde{b}_m} c_m(t),
\end{equation} 
where $\Omega^{-1}$ is the inverse of the overlap matrix
$\Omega_{jl} = \braket{\tilde{b}_j}{\tilde{b}_l} = (S^{-1})_{jl}$ 
of the bvN basis,
and $H(t)$ is the Hamiltonian operator of the system. 
The overlap and Hamiltonian matrix elements in eq. (\ref{eq: TDSE}) 
can be computed simply via the representations in 
$\{\ket{\theta_m}\}$ and 
$\{\ket{\phi_m}\}$. \cite{jcp91_3571,arXiv1201_2299v1}
The matrix $\Omega^{-1}$ is Hermitian positive-definite, 
and the elements $\{\bra{\tilde{b}_l} H(t) \ket{\tilde{b}_m}\}$
constitute an  Hermitian matrix.
Therefore, the product of these matrices yields
all real eigenvalues \cite{Bai_etal2000}, 
and eq. (\ref{eq: TDSE}) can be solved stably 
by many standard numerical algorithms.

\begin{figure}[bt]
\begin{center}
\includegraphics[width=\linewidth]{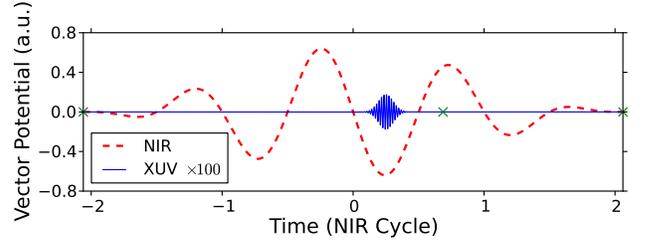}
\caption{Vector potentials, $A_{\text{NIR}}(t)$ and $A_{\text{XUV}}(t)$,
of the NIR and XUV laser pulses applied to the model 1D atom. 
}
\label{fig: vecpot}
\end{center}
\end{figure}

To demonstrate the accuracy and efficiency of the present method,
we solve eq. (\ref{eq: TDSE}) for the electronic wavepacket of a 1D atom 
in the combined field of NIR and XUV laser pulses.
The Hamiltonian of this system is given as
\begin{equation}\label{eq: Ht}
H(t) = H_0 + V(t),
\end{equation}
where $H_0$ is the field-free Hamiltonian expressed as 
\begin{equation}
H_0 = \frac{p^2}{2\mu} 
     -\frac{Q e^2}{4\pi\epsilon_0\sqrt{x^2+a^2}}. 
\end{equation}
Here $\mu = 1$ a.u. is the electron mass, $e= -1$ a.u. is the electron 
charge, $-Qe=1$ a.u. is the charge of the atomic nucleus, 
$a=1$ a.u. is the soft-core parameter,  
and $\epsilon_0=1/4\pi$ a.u. is the electric constant.
The laser-electron coupling $V(t)$ is, in the velocity gauge,
\begin{equation}
V(t)= -\frac{e}{\mu} [A_{\text{NIR}}(t) + A_{\text{XUV}}(t)]p, 
\end{equation}
where $A_{\text{NIR}}(t)$ and $A_{\text{XUV}}(t)$ are
the vector potentials of the NIR and XUV laser pulses, respectively. 
We used an NIR pulse of wavelength $800$ nm, 
peak intensity $5\times 10^{13}$ W/cm$^2$, and duration $1.5$ cycles
(4 fs, FWHM of intensity profile).
The XUV pulse had wavelength $15$ nm, 
peak intensity $1\times 10^{12}$ W/cm$^2$, 
and duration $5.0$ cycles (250 as). 
The peak of the XUV pulse was delayed from that of the NIR pulse by
$0.25$ NIR cycles, as shown in Fig. \ref{fig: vecpot}.


The initial state was chosen as the ground state of 
$H_0$ with energy eigenvalue $-0.66978$ a.u.,
and the wavepacket was propagated 
from the turn-on of the NIR laser pulse at $t_{\text{min}}$ 
to its end at $t_{\text{max}}$
by the short-iterative Arnoldi algorithm \cite{jcp100_5054,jcp114_1497}
in a $6$D Krylov space 
with a constant time-step of $\Delta t=0.0379$ a.u.
We divided the time span from $t_{\text{min}}$ to $t_{\text{max}}$
into $8$ time segments and changed the active
set $\mathcal{A}$ from one segment to the next. 
The number of FG points was $N=4096$, and they were distributed
over $-750$ a.u. $\leq x \leq 750$ a.u. 
and $-8.58$ a.u. $\leq p \leq 8.58$ a.u.
This phase space rectangle was divided into $64 \times 64$ cells of aspect ratio 
$\hbar\gamma=1.14\times 10^{-2}$ a.u., each of which contained a Gaussian center
$(q_j,p_j)$.


\begin{figure}[btp]
\begin{center}
\includegraphics[width=.90\linewidth]{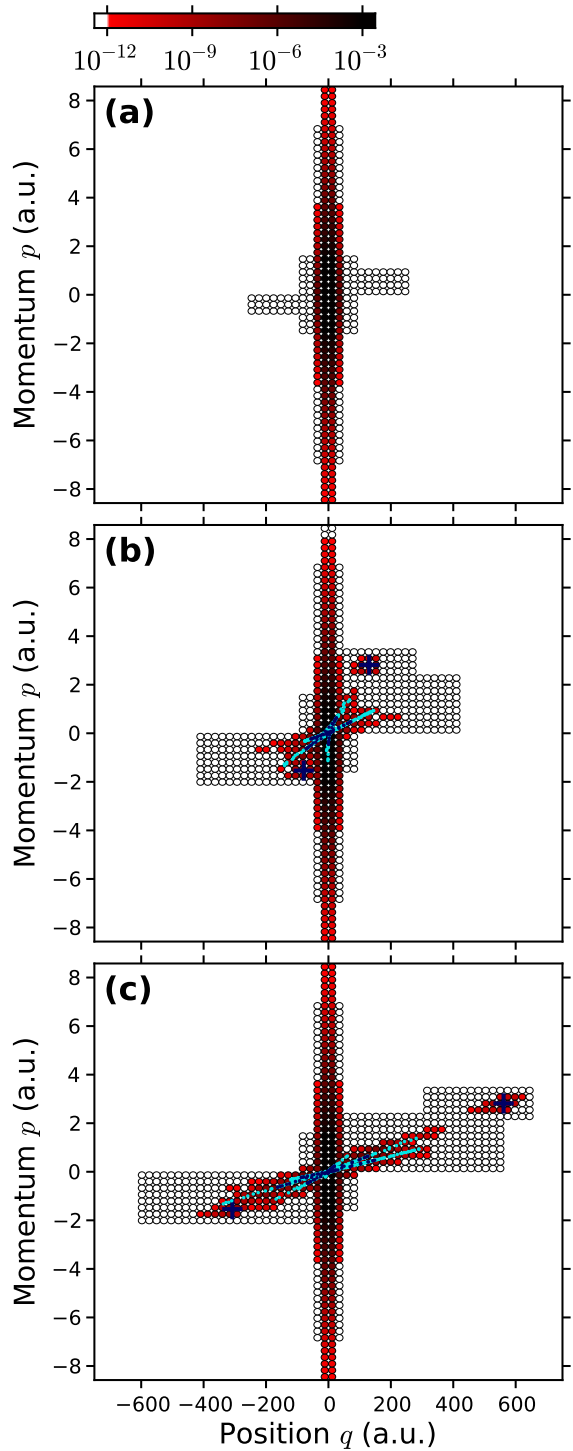}
\caption{Snapshots of $\{|c_j|^2\}_{j\in \mathcal{A}}$
shown by the ellipses 
located at the Gaussian centers $\{(q_j, p_j)\}_{j\in \mathcal{A}}$.  
The colors of the ellipses indicate the magnitude of $|c_j|^2$
according to the scale above the figure. 
The sequence of dark blue dots represent the simple-man trajectories
for direct ionization; the light blue dots represent the rescattered simple-man
trajectories.
The dark blue $+$ marks represent the simple-man 
trajectories absorbing one XUV photon in the presence of the NIR field. 
The snapshots were taken at (a) $t= -2.06$,
(b) $t=0.69$, and (c) $t=2.06$ in units of NIR cycles. 
These times are indicated by 
the green $\times$ marks in Fig. \ref{fig: vecpot}.
}
\label{fig: snap_qp}
\end{center}
\end{figure}

In Fig. \ref{fig: snap_qp}, snapshots of 
$|c_j(t)|^2$ are 
shown by the color scale of the ellipses located at 
the active Gaussian centers $\{(q_j, p_j)\}_{j\in \mathcal{A}}$.
The outer rectangular boundary of each panel 
indicates the phase space area corresponding to
the Hilbert space $\mathcal{H}$ spanned respectively
by the Fourier spectral and pseudospectral bases as well
as the full pvN and the full bvN bases.
The wavepacket, initially concentrated at the atomic 
core [Fig. \ref{fig: snap_qp}(a)], is ionized by the 
NIR and XUV laser pulses and spreads into parts
of the first ($x>0$ and $p>0$) and third ($x<0$ and $p<0$)
quadrants [Fig. \ref{fig: snap_qp}(b,c)], 
but a large area is never accessed.
This can be intuitively expected from the classical 
mechanics, and indeed we see that
the wavepacket closely follows the
so-called simple-man trajectories \cite{prl71_1994}
(dots and $+$ marks in Fig. \ref{fig: snap_qp})
which obey the classical Hamiltonian of the same form as
eq. (\ref{eq: Ht}) with $H_0$ replaced by $p^2/2\mu$.
In fact, we chose $\mathcal{A}$ so that
the active phase space domain contains 
these simple-man trajectories (with an additional margin).

%

\begin{figure}[btp]
\begin{center}
\includegraphics[width=\linewidth]{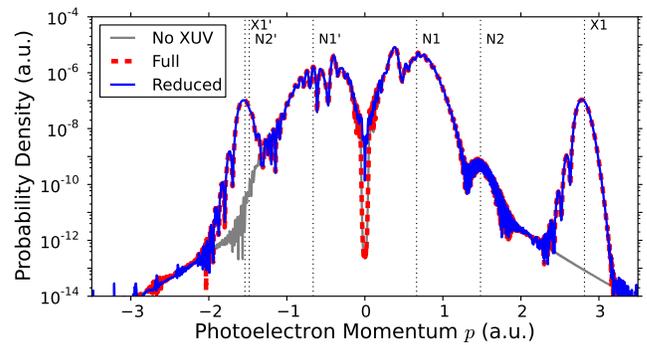}
\caption{
Comparison of the 
photoelectron momentum distributions obtained with the 
reduced bvN basis (blue solid line) and 
full bvN basis (red dashed line).
The momentum distribution from a simulation
without the XUV pulse (using the full bvN basis) is also shown 
(gray solid line).
The vertical dashed lines indicate
the cut-offs of the direct (N1 and N1')
and rescattered (N2 and N2') photoelectrons, as well as
the NIR-streaked single-XUV-photon ionization peaks (X1 and X1'),
estimated by the simple-man model.
}
\label{fig: photomom}
\end{center}
\end{figure}

In Fig. \ref{fig: photomom}, we 
compare the photoelectron momentum distributions
obtained using the reduced basis of Fig. \ref{fig: snap_qp} 
and the full bvN basis.
The excellent agreement between the two results
indicates that the present method not only 
preserves the qualitative features --- cut-offs of 
the direct and rescattered NIR photoelectrons, and 
the NIR-streaked single-XUV-photon 
ionization peaks --- but also has quantitative accuracy.

\begin{figure}[btp]
\begin{center}
\includegraphics[width=\linewidth]{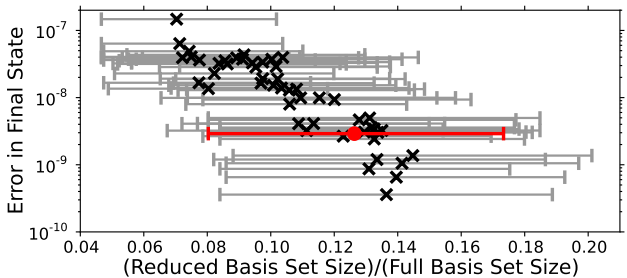}
\caption{The error $\epsilon$  
as a function of $\langle N_{\mathcal{A}} \rangle/N$ (black $\times$ marks).
The horizontal error bars indicate
the range of $N_{\mathcal{A}}(t)/N$
in $t_{\text{min}} \leq t \leq t_{\text{max}}$.
The data marked by the red filled circle is
from the simulation shown in
Figs. \ref{fig: snap_qp} and \ref{fig: photomom}.
}
\label{fig: accuracy_vs_basissize}
\end{center}
\end{figure}

The accuracy and the size of the active bvN basis set are 
expected to be inversely related.
However, it is not straightforward to determine 
the minimal size of the active set that maintains a given accuracy. 
To seek an upper bound to such an optimal basis size, 
we carried out simulations with different active sets
generated by changing the margins around 
the simple-man trajectories.
The error of a simulation was measured by
$\epsilon :=
|\braket{\Psi^{\text{reduc}}(t_{\text{max}})}{\Psi^{\text{full}}(t_{\text{max}})}-1|$,
where $\ket{\Psi^{\text{reduc}}(t_{\text{max}})}$ 
and $\ket{\Psi^{\text{full}}(t_{\text{max}})}$ are the final states
calculated using a reduced and the full bvN bases, respectively.
Figure \ref{fig: accuracy_vs_basissize} shows the dependence
of this error on the basis set size.
As the size $N_{\mathcal{A}}$ is time-dependent, 
and as the computational cost of wavepacket propagation 
depends on  $N_{\mathcal{A}}$ quadratically, 
we characterized  $N_{\mathcal{A}}$ by its root-mean-square,
$\langle N_{\mathcal{A}} \rangle
  := \sqrt{\int_{t_{\text{min}}}^{t_{\text{max}}} dt
     [N_{\mathcal{A}}(t)]^2/(t_{\text{max}}-t_{\text{min}})}$,
as well as its minimum and maximum values.
It can be seen
that the bvN basis can be compressed down to 
$\langle N_{\mathcal{A}} \rangle/N = 0.14 $ or possibly further
while maintaining the accuracy level at $\epsilon = 4\times 10^{-10}$, 
and at least to
$\langle N_{\mathcal{A}} \rangle/N = 0.08$ 
for $\epsilon = 6\times 10^{-8}$.

The computational cost of the present method is dominated
by the multiplication of the matrix
$G_{jm} := (-i/\hbar) \sum_{l \in \mathcal{A}} (\Omega^{-1})_{j l} 
  \bra{\tilde{b}_l} H(t) \ket{\tilde{b}_m}$
and the vector $c_m$ ($j,m\in \mathcal{A}$) 
which scales as $O(N_{\mathcal{A}}^2)$.
There is also initial overhead that scales as $O(N^3)$
originating from the computation of 
$S_{mj}$ and $(S^{-1})_{jl}$ ($m,j=1,\dots,N$; $l \in \mathcal{A}$).

Our method can be extended straightforwardly to systems with multiple 
DOF. \cite{arXiv1201_2299v1}
Defining
the number $d$ of DOF, the total number $M$ of FG points increases
exponentially as $M=O(N^d)$. In contrast, the number $M_{\mathcal{A}}$ of 
active bvN basis states 
for atomic and molecular systems in a laser field
is expected to scale more slowly than $M$
since the phase space coordinates are generally correlated
for such systems. \cite{prl99_263002,prl107_143004,jcp89_6860}
The cost per time step of the present method goes as $O(M_{\mathcal{A}}^2)$.
The initial overhead in 
calculating $S_{mj}$ and $(S^{-1})_{jl}$
stays at $O(N^3)$ because the multi-D Gaussians
can be factored into 1D Gaussians \cite{arXiv1201_2299v1}.
The dominant part in the overhead is now the calculation of the 
matrix elements for the potential energy operator in the reduced 
bvN basis, and this goes as $O(M M_{\mathcal{A}}^2)$.
For sufficiently large $d$,
the present method has the potential to perform better than the popular
alternate-direction Crank-Nicolson 
scheme \cite{procCambPhilSoc43_50,jSocIndustApplMath3_28} 
[which costs $O(M)$ per step]
and the FG split-operator method \cite{jCompPhys47_412,jPhysChem92_2087}
[which costs $O(M \log_2 N)$ per step]. 
The bvN basis may also
be used for the one-body orbitals in MCTDHF as an alternative 
to the static scaling of the position coordinates
\cite{jCompPhys130_148,jcp111_9498,jcp113_8953} 
used previously. \cite{jcp128_184102}


In summary, we presented a new method to solve the TDSE based on the bvN 
basis. Although the basis is time-independent,
the active subset is chosen in a time-dependent manner. 
As a first demonstration, we calculated 
the electronic wavepacket of a 1D atom 
in the combined fields of intense NIR and attosecond XUV laser pulses.
This example demonstrates the 
high accuracy and efficiency of the method. 
We are currently working to extend the method to 3D with the 
aim of ultimately applying it to
multi-electron systems in intense and ultrashort laser pulses.

This work was supported by the Minerva Foundation
and the Israel Science Foundation under Grant No.\ 807/08. 
This research is made possible by the historic generosity of the
Harold Perlman family.

%


\begin{thebibliography}{10}%
\makeatletter
\providecommand \@ifxundefined [1]{%
 \ifx #1\undefined \expandafter \@firstoftwo
 \else \expandafter \@secondoftwo
\fi
}%
\providecommand \@ifnum [1]{%
 \ifnum #1\expandafter \@firstoftwo
 \else \expandafter \@secondoftwo
\fi
}%
\providecommand \enquote [1]{``#1''}%
\providecommand \bibnamefont  [1]{#1}%
\providecommand \bibfnamefont [1]{#1}%
\providecommand \citenamefont [1]{#1}%
\providecommand\href[0]{\@sanitize\@href}%
\providecommand\@href[1]{\endgroup\@@startlink{#1}\endgroup\@@href}%
\providecommand\@@href[1]{#1\@@endlink}%
\providecommand \@sanitize [0]{\begingroup\catcode`\&12\catcode`\#12\relax}%
\@ifxundefined \pdfoutput {\@firstoftwo}{%
 \@ifnum{\z@=\pdfoutput}{\@firstoftwo}{\@secondoftwo}%
}{%
 \providecommand\@@startlink[1]{\leavevmode}%
 \providecommand\@@endlink[0]{}%
}{%
 \providecommand\@@startlink[1]{%
  \leavevmode
  \pdfstartlink
   attr{/Border[0 0 1 ]/H/I/C[0 1 1]}%
   user{/Subtype/Link/A<</Type/Action/S/URI/URI(#1)>>}%
  \relax
 }%
 \providecommand\@@endlink[0]{\pdfendlink}%
}%
\providecommand \url  [0]{\begingroup\@sanitize \@url }%
\providecommand \@url [1]{\endgroup\@href {#1}{\urlprefix}}%
\providecommand \urlprefix [0]{URL }%
\providecommand \Eprint[0]{\href }%
\@ifxundefined \urlstyle {%
  \providecommand \doi [1]{doi:\discretionary{}{}{}#1}%
}{%
  \providecommand \doi [0]{doi:\discretionary{}{}{}\begingroup
  \urlstyle{rm}\Url }%
}%
\providecommand \doibase [0]{http://dx.doi.org/}%
\providecommand \Doi[1]{\href{\doibase#1}}%
\providecommand \selectlanguage [0]{\@gobble}%
\providecommand \bibinfo [0]{\@secondoftwo}%
\providecommand \bibfield [0]{\@secondoftwo}%
\providecommand \translation [1]{[#1]}%
\providecommand \BibitemOpen[0]{}%
\providecommand \bibitemStop [0]{}%
\providecommand \bibitemNoStop [0]{.\EOS\space}%
\providecommand \EOS [0]{\spacefactor3000\relax}%
\providecommand \BibitemShut [1]{\csname bibitem#1\endcsname}%
\bibitem{naturePhotonics4_822}%
  \BibitemOpen
  \bibfield{author}{%
  \bibinfo {author} {\bibfnamefont{T.}~\bibnamefont{Popmintchev}}
  \emph{et~al.},\ }%
  \bibfield{journal}{%
  \bibinfo {journal} {Nature Photonics}\ }%
  \textbf{\bibinfo {volume} {4}},\ \bibinfo {pages} {822} (\bibinfo {year}
  {2010})\BibitemShut{NoStop}%
\bibitem{revModPhys81_163}%
  \BibitemOpen
  \bibfield{author}{%
  \bibinfo {author} {\bibfnamefont{F.}~\bibnamefont{Krausz}}\ and\ \bibinfo
  {author} {\bibfnamefont{M.}~\bibnamefont{Ivanov}},\ }%
  \bibfield{journal}{%
  \bibinfo {journal} {Rev. Mod. Phys.}\ }%
  \textbf{\bibinfo {volume} {81}},\ \bibinfo {pages} {163} (\bibinfo {year}
  {2009})\BibitemShut{NoStop}%
\bibitem{prl105_203004}%
  \BibitemOpen
  \bibfield{author}{%
  \bibinfo {author} {\bibfnamefont{N.}~\bibnamefont{Takemoto}}\ and\ \bibinfo
  {author} {\bibfnamefont{A.}~\bibnamefont{Becker}},\ }%
  \bibfield{journal}{%
  \bibinfo {journal} {Phys. Rev. Lett.}\ }%
  \textbf{\bibinfo {volume} {105}},\ \bibinfo {pages} {203004} (\bibinfo {year}
  {2010})\BibitemShut{NoStop}%
\bibitem{prl107_143004}%
  \BibitemOpen
  \bibfield{author}{%
  \bibinfo {author} {\bibfnamefont{M.}~\bibnamefont{Odenweller}}
  \emph{et~al.},\ }%
  \bibfield{journal}{%
  \bibinfo {journal} {Phys. Rev. Lett.}\ }%
  \textbf{\bibinfo {volume} {107}},\ \bibinfo {pages} {143004} (\bibinfo {year}
  {2011})\BibitemShut{NoStop}%
\bibitem{prl71_1994}%
  \BibitemOpen
  \bibfield{author}{%
  \bibinfo {author} {\bibfnamefont{P.~B.}\ \bibnamefont{Corkum}},\ }%
  \bibfield{journal}{%
  \bibinfo {journal} {Phys. Rev. Lett.}\ }%
  \textbf{\bibinfo {volume} {71}},\ \bibinfo {pages} {1994} (\bibinfo {year}
  {1993})\BibitemShut{NoStop}%
\bibitem{newJPhys10_025019}%
  \BibitemOpen
  \bibfield{author}{%
  \bibinfo {author} {\bibfnamefont{M.}~\bibnamefont{Nest}}, \bibinfo {author}
  {\bibfnamefont{F.}~\bibnamefont{Remacle}},\ and\ \bibinfo {author}
  {\bibfnamefont{R.~D.}\ \bibnamefont{Levine}},\ }%
  \bibfield{journal}{%
  \bibinfo {journal} {New J. Phys.}\ }%
  \textbf{\bibinfo {volume} {10}},\ \bibinfo {pages} {025019} (\bibinfo {year}
  {2008})\BibitemShut{NoStop}%
\bibitem{jPhysB36_L393}%
  \BibitemOpen
  \bibfield{author}{%
  \bibinfo {author} {\bibfnamefont{J.~S.}\ \bibnamefont{Parker}}
  \emph{et~al.},\ }%
  \bibfield{journal}{%
  \bibinfo {journal} {J. Phys. B}\ }%
  \textbf{\bibinfo {volume} {36}},\ \bibinfo {pages} {L393} (\bibinfo {year}
  {2003})\BibitemShut{NoStop}%
\bibitem{jcp113_8953}%
  \BibitemOpen
  \bibfield{author}{%
  \bibinfo {author} {\bibfnamefont{K.}~\bibnamefont{Harumiya}} \emph{et~al.},\
  }%
  \bibfield{journal}{%
  \bibinfo {journal} {J. Chem. Phys.}\ }%
  \textbf{\bibinfo {volume} {113}},\ \bibinfo {pages} {8953} (\bibinfo {year}
  {2000})\BibitemShut{NoStop}%
\bibitem{laserPhys13_1064}%
  \BibitemOpen
  \bibfield{author}{%
  \bibinfo {author} {\bibfnamefont{J.}~\bibnamefont{Zanghellini}}
  \emph{et~al.},\ }%
  \bibfield{journal}{%
  \bibinfo {journal} {Laser Phys.}\ }%
  \textbf{\bibinfo {volume} {13}},\ \bibinfo {pages} {1064} (\bibinfo {year}
  {2003})\BibitemShut{NoStop}%
\bibitem{jcp128_184102}%
  \BibitemOpen
  \bibfield{author}{%
  \bibinfo {author} {\bibfnamefont{T.}~\bibnamefont{Kato}}\ and\ \bibinfo
  {author} {\bibfnamefont{H.}~\bibnamefont{Kono}},\ }%
  \bibfield{journal}{%
  \bibinfo {journal} {J. Chem. Phys.}\ }%
  \textbf{\bibinfo {volume} {128}},\ \bibinfo {pages} {184102} (\bibinfo {year}
  {2008})\BibitemShut{NoStop}%
\bibitem{prl96_073004}%
  \BibitemOpen
  \bibfield{author}{%
  \bibinfo {author} {\bibfnamefont{S.~X.}\ \bibnamefont{Hu}}\ and\ \bibinfo
  {author} {\bibfnamefont{L.~A.}\ \bibnamefont{Collins}},\ }%
  \bibfield{journal}{%
  \bibinfo {journal} {Phys. Rev. Lett.}\ }%
  \textbf{\bibinfo {volume} {96}},\ \bibinfo {pages} {073004} (\bibinfo {year}
  {2006})\BibitemShut{NoStop}%
\bibitem{physRevA83_063416}%
  \BibitemOpen
  \bibfield{author}{%
  \bibinfo {author} {\bibfnamefont{D.~J.}\ \bibnamefont{Haxton}}, \bibinfo
  {author} {\bibfnamefont{K.~V.}\ \bibnamefont{Lawler}},\ and\ \bibinfo
  {author} {\bibfnamefont{C.~W.}\ \bibnamefont{McCurdy}},\ }%
  \bibfield{journal}{%
  \bibinfo {journal} {Phys. Rev. A}\ }%
  \textbf{\bibinfo {volume} {83}},\ \bibinfo {pages} {063416} (\bibinfo {year}
  {2011})\BibitemShut{NoStop}%
\bibitem{Low1985}%
  \BibitemOpen
  \bibfield{author}{%
  \bibinfo {author} {\bibfnamefont{F.}~\bibnamefont{Low}},\ }%
  in\ \emph{\bibinfo {booktitle} {A Passion for Physics -- Essays in Honor of
  Geoffrey Chew}}\ (\bibinfo {publisher} {World Scientific},\ \bibinfo {year}
  {1985})\ pp.\ \bibinfo {pages} {17--22}\BibitemShut{NoStop}%
\bibitem{jcp113_10028}%
  \BibitemOpen
  \bibfield{author}{%
  \bibinfo {author} {\bibfnamefont{D.~V.}\ \bibnamefont{Shalashilin}}\ and\
  \bibinfo {author} {\bibfnamefont{M.~S.}\ \bibnamefont{Child}},\ }%
  \bibfield{journal}{%
  \bibinfo {journal} {J. Chem. Phys.}\ }%
  \textbf{\bibinfo {volume} {113}},\ \bibinfo {pages} {10028} (\bibinfo {year}
  {2000})\BibitemShut{NoStop}%
\bibitem{chemPhys304_103}%
  \BibitemOpen
  \bibfield{author}{%
  \bibinfo {author} {\bibfnamefont{D.~V.}\ \bibnamefont{Shalashilin}}\ and\
  \bibinfo {author} {\bibfnamefont{M.~S.}\ \bibnamefont{Child}},\ }%
  \bibfield{journal}{%
  \bibinfo {journal} {Chem. Phys.}\ }%
  \textbf{\bibinfo {volume} {304}},\ \bibinfo {pages} {103} (\bibinfo {year}
  {2004})\BibitemShut{NoStop}%
\bibitem{jcp115_1158}%
  \BibitemOpen
  \bibfield{author}{%
  \bibinfo {author} {\bibfnamefont{L.}~\bibnamefont{{Mauritz Andersson}}},\ }%
  \bibfield{journal}{%
  \bibinfo {journal} {J. Chem. Phys.}\ }%
  \textbf{\bibinfo {volume} {115}},\ \bibinfo {pages} {1158} (\bibinfo {year}
  {2001})\BibitemShut{NoStop}%
\bibitem{jcp118_2606}%
  \BibitemOpen
  \bibfield{author}{%
  \bibinfo {author} {\bibfnamefont{M.}~\bibnamefont{Satta}}, \bibinfo {author}
  {\bibfnamefont{E.}~\bibnamefont{Scifoni}},\ and\ \bibinfo {author}
  {\bibfnamefont{F.~A.}\ \bibnamefont{Gianturco}},\ }%
  \bibfield{journal}{%
  \bibinfo {journal} {J. Chem. Phys.}\ }%
  \textbf{\bibinfo {volume} {118}},\ \bibinfo {pages} {2606} (\bibinfo {year}
  {2003})\BibitemShut{NoStop}%
\bibitem{jcp124_204101}%
  \BibitemOpen
  \bibfield{author}{%
  \bibinfo {author} {\bibfnamefont{D.~A.}\ \bibnamefont{McCormack}},\ }%
  \bibfield{journal}{%
  \bibinfo {journal} {J. Chem. Phys.}\ }%
  \textbf{\bibinfo {volume} {124}},\ \bibinfo {pages} {204101} (\bibinfo {year}
  {2006})\BibitemShut{NoStop}%
\bibitem{jcp118_6720}%
  \BibitemOpen
  \bibfield{author}{%
  \bibinfo {author} {\bibfnamefont{Y.}~\bibnamefont{Wu}}\ and\ \bibinfo
  {author} {\bibfnamefont{V.~S.}\ \bibnamefont{Batista}},\ }%
  \bibfield{journal}{%
  \bibinfo {journal} {J. Chem. Phys.}\ }%
  \textbf{\bibinfo {volume} {118}},\ \bibinfo {pages} {6720} (\bibinfo {year}
  {2003})\BibitemShut{NoStop}%
\bibitem{newJPhys11_105052}%
  \BibitemOpen
  \bibfield{author}{%
  \bibinfo {author} {\bibfnamefont{F.}~\bibnamefont{Dimler}} \emph{et~al.},\ }%
  \bibfield{journal}{%
  \bibinfo {journal} {New J. Phys.}\ }%
  \textbf{\bibinfo {volume} {11}},\ \bibinfo {pages} {105052} (\bibinfo {year}
  {2009})\BibitemShut{NoStop}%
\bibitem{arXiv1201_2299v1}%
  \BibitemOpen
  \bibfield{author}{%
  \bibinfo {author} {\bibfnamefont{A.}~\bibnamefont{Shimshovitz}}\ and\
  \bibinfo {author} {\bibfnamefont{D.~J.}\ \bibnamefont{Tannor}},\ }%
  \bibfield{journal}{%
  \bibinfo {journal} {arXiv}\ }%
  \textbf{\bibinfo {volume} {1201}},\ \bibinfo {pages} {2299v1} (\bibinfo
  {year} {2012})\BibitemShut{NoStop}%
\bibitem{Kosloff1993}%
  \BibitemOpen
  \bibfield{author}{%
  \bibinfo {author} {\bibfnamefont{R.}~\bibnamefont{Kosloff}},\ }%
  in\ \emph{\bibinfo {booktitle} {Numerical Grid Methods and Their Application
  to Schr{\"{o}}dinger Equation}},\ \bibinfo {editor} {edited by\ \bibinfo
  {editor} {\bibfnamefont{C.}~\bibnamefont{Cerjan}}}\ (\bibinfo {publisher}
  {Kluwer},\ \bibinfo {year} {1993})\ pp.\ \bibinfo {pages}
  {175--194}\BibitemShut{NoStop}%
\bibitem{jcp91_3571}%
  \BibitemOpen
  \bibfield{author}{%
  \bibinfo {author} {\bibfnamefont{C.~C.}\ \bibnamefont{Marston}}\ and\
  \bibinfo {author} {\bibfnamefont{G.~G.}\ \bibnamefont{Balint-Kurti}},\ }%
  \bibfield{journal}{%
  \bibinfo {journal} {J. Chem. Phys.}\ }%
  \textbf{\bibinfo {volume} {91}},\ \bibinfo {pages} {3571} (\bibinfo {year}
  {1989})\BibitemShut{NoStop}%
\bibitem{jCompPhys47_412}%
  \BibitemOpen
  \bibfield{author}{%
  \bibinfo {author} {\bibfnamefont{M.}~\bibnamefont{Feit}}, \bibinfo {author}
  {\bibfnamefont{J.}~\bibnamefont{{Fleck Jr.}}},\ and\ \bibinfo {author}
  {\bibfnamefont{A.}~\bibnamefont{Steiger}},\ }%
  \bibfield{journal}{%
  \bibinfo {journal} {J. Comp. Phys.}\ }%
  \textbf{\bibinfo {volume} {47}},\ \bibinfo {pages} {412} (\bibinfo {year}
  {1982})\BibitemShut{NoStop}%
\bibitem{jPhysChem92_2087}%
  \BibitemOpen
  \bibfield{author}{%
  \bibinfo {author} {\bibfnamefont{R.}~\bibnamefont{Kosloff}},\ }%
  \bibfield{journal}{%
  \bibinfo {journal} {J. Phys. Chem.}\ }%
  \textbf{\bibinfo {volume} {92}},\ \bibinfo {pages} {2087} (\bibinfo {year}
  {1988})\BibitemShut{NoStop}%
\bibitem{Bai_etal2000}%
  \BibitemOpen
  \emph{\bibinfo {title} {Templates for the Solution of Algebraic Eigenvalue
  Problems}},\ edited by\ \bibinfo {editor}
  {\bibfnamefont{Z.}~\bibnamefont{Bai}}, \bibinfo {editor}
  {\bibfnamefont{J.}~\bibnamefont{Demmel}}, \bibinfo {editor}
  {\bibfnamefont{J.}~\bibnamefont{Dongarra}}, \bibinfo {editor}
  {\bibfnamefont{A.}~\bibnamefont{Ruhe}},\ and\ \bibinfo {editor}
  {\bibfnamefont{H.}~\bibnamefont{van~der Vorst}}\ (\bibinfo {publisher}
  {SIAM},\ \bibinfo {address} {Philadelphia},\ \bibinfo {year} {2000})\
  \bibinfo {note} {{Ch. 5}}\BibitemShut{NoStop}%
\bibitem{jcp100_5054}%
  \BibitemOpen
  \bibfield{author}{%
  \bibinfo {author} {\bibfnamefont{W.~T.}\ \bibnamefont{Pollard}}\ and\
  \bibinfo {author} {\bibfnamefont{R.~A.}\ \bibnamefont{Friesner}},\ }%
  \bibfield{journal}{%
  \bibinfo {journal} {J. Chem. Phys.}\ }%
  \textbf{\bibinfo {volume} {100}},\ \bibinfo {pages} {5054} (\bibinfo {year}
  {1994})\BibitemShut{NoStop}%
\bibitem{jcp114_1497}%
  \BibitemOpen
  \bibfield{author}{%
  \bibinfo {author} {\bibfnamefont{I.}~\bibnamefont{Kondov}}, \bibinfo {author}
  {\bibfnamefont{U.}~\bibnamefont{Kleinekath{\"{o}}fer}},\ and\ \bibinfo
  {author} {\bibfnamefont{M.}~\bibnamefont{Schreiber}},\ }%
  \bibfield{journal}{%
  \bibinfo {journal} {J. Chem. Phys.}\ }%
  \textbf{\bibinfo {volume} {114}},\ \bibinfo {pages} {1497} (\bibinfo {year}
  {2001})\BibitemShut{NoStop}%
\bibitem{prl99_263002}%
  \BibitemOpen
  \bibfield{author}{%
  \bibinfo {author} {\bibfnamefont{A.}~\bibnamefont{Staudte}} \emph{et~al.},\
  }%
  \bibfield{journal}{%
  \bibinfo {journal} {Phys. Rev. Lett.}\ }%
  \textbf{\bibinfo {volume} {99}},\ \bibinfo {pages} {263002} (\bibinfo {year}
  {2007})\BibitemShut{NoStop}%
\bibitem{jcp89_6860}%
  \BibitemOpen
  \bibfield{author}{%
  \bibinfo {author} {\bibfnamefont{J.~L.}\ \bibnamefont{Anchell}}\ and\
  \bibinfo {author} {\bibfnamefont{J.~E.}\ \bibnamefont{Harriman}},\ }%
  \bibfield{journal}{%
  \bibinfo {journal} {J. Chem. Phys.}\ }%
  \textbf{\bibinfo {volume} {89}},\ \bibinfo {pages} {6860} (\bibinfo {year}
  {1988})\BibitemShut{NoStop}%
\bibitem{procCambPhilSoc43_50}%
  \BibitemOpen
  \bibfield{author}{%
  \bibinfo {author} {\bibfnamefont{J.}~\bibnamefont{Crank}}\ and\ \bibinfo
  {author} {\bibfnamefont{P.}~\bibnamefont{Nicolson}},\ }%
  \bibfield{journal}{%
  \bibinfo {journal} {Proc. Camb. Phil. Soc.}\ }%
  \textbf{\bibinfo {volume} {43}},\ \bibinfo {pages} {50} (\bibinfo {year}
  {1947})\BibitemShut{NoStop}%
\bibitem{jSocIndustApplMath3_28}%
  \BibitemOpen
  \bibfield{author}{%
  \bibinfo {author} {\bibfnamefont{D.~W.}\ \bibnamefont{Peaceman}}\ and\
  \bibinfo {author} {\bibfnamefont{J.}~\bibnamefont{H.~H.~Rachford}},\ }%
  \bibfield{journal}{%
  \bibinfo {journal} {J. Soc. Indust. Appl. Math.}\ }%
  \textbf{\bibinfo {volume} {3}},\ \bibinfo {pages} {28} (\bibinfo {year}
  {1955})\BibitemShut{NoStop}%
\bibitem{jCompPhys130_148}%
  \BibitemOpen
  \bibfield{author}{%
  \bibinfo {author} {\bibfnamefont{H.}~\bibnamefont{Kono}} \emph{et~al.},\ }%
  \bibfield{journal}{%
  \bibinfo {journal} {J. Comp. Phys.}\ }%
  \textbf{\bibinfo {volume} {130}},\ \bibinfo {pages} {148} (\bibinfo {year}
  {1997})\BibitemShut{NoStop}%
\bibitem{jcp111_9498}%
  \BibitemOpen
  \bibfield{author}{%
  \bibinfo {author} {\bibfnamefont{I.}~\bibnamefont{Kawata}}\ and\ \bibinfo
  {author} {\bibfnamefont{H.}~\bibnamefont{Kono}},\ }%
  \bibfield{journal}{%
  \bibinfo {journal} {J. Chem. Phys.}\ }%
  \textbf{\bibinfo {volume} {111}},\ \bibinfo {pages} {9498} (\bibinfo {year}
  {1999})\BibitemShut{NoStop}%
\end{thebibliography}

%

\end{document}